# Friction, Free Axes of Rotation and Entropy[1]

Alexander Kazachkov[1], Victor Multanen[2], Viktor Danchuk[3]; and Mark Frenkel[3], Edward Bormashenko[2,*]

1. V. N. Karazin Kharkiv National University, Ukraine, School of Physics, Svobody Sq. 4, 61022, Kharkiv, Ukraine
2. Ariel University, Engineering Sciences Faculty, Department of Chemical Technology, 407000, P.O.B. 3, Ariel, Israel
3. Ariel University, Exact Sciences Faculty, Department of Physics, 407000, P.O.B. 3, Ariel, Israel
* Correspondence: edward@ariel.ac.il

**Abstract:** Friction forces acting on rotators may promote their alignment and therefore eliminate degrees of freedom in their movement. The alignment of rotators by friction force was demonstrated by experiments performed with different spinners, demonstrating how friction generates negentropy in a system of rotators. A gas of rigid rotators influenced by friction force is considered. The orientational negentropy generated by a friction force was estimated with the Sackur-Tetrode equation. The minimal change in total entropy of a system of rotators, corresponding to their eventual alignment, decreases with temperature. The reported effect may be of the primary importance for a phase equilibrium and motion of ubiquitous colloidal and granular systems.

## 1. Introduction

Interrelations between a friction and entropy are perplexing and ambiguous. Friction forces usually result in irreversibility and generation of entropy [1-4]. However, as it was already demonstrated by Nosonovsky et al., friction force may lead to self-organization, patterning and production of negentropy [5-9]. The pattern formation may involve self-organized criticality and reaction–diffusion systems [6].The self-organization and patterning inspired by friction forces give rise to rich physical content in multi-scaled, hierarchical systems [7]. In certain cases, the entropy production at a particular scale level may be compensated for by the entropy consumption at another level [7]. Possible mechanisms of the decrease of entropy under dry friction have been discussed in Ref. 8. From the practical point of view, the friction-induced self-organization is important due to the self-healing and self-lubrication arising from negentropy generation, which is inherent for these phenomena [9].

Our paper is devoted to the friction-induced orientation of rotating rigid spinners, resulting in the decrease of the entropy of a gas containing such spinners; in other words, in the production of orientational negentropy. The effects related to orientational entropy have already been extensively addressed in the context of phase transitions in polymers and liquid crystals [10-11]. Entropic contributions to common ordering transitions are essential for the design of self-assembling systems with addressable complexity [12]. However, the friction-induced generation of negentropy, occurring under re-orientation of rotating bodies, is first discussed in this paper. The reported effect may be essential for a phase equilibrium and motion of colloidal and granular systems, widespread in nature and engineering.

---

## 2. Friction induced orientation of a rotator.

The friction-induced orientation of a rotating rigid body is demonstrated in the following series of experiments.

The first series of experiments was performed with the hollow top depicted in **Figure 1A**. The mass of the top was varied from 8.650 to 99.330 grams by filling its conical cavity with steel balls, as shown in Figure 1B. The mass and diameter of the balls were $m=0.465\pm0.0005$g and $D=4.8\pm0.05$mm. The angular velocities of the full and hollow tops were $\omega_f = 25\pm0.5 s^{-1}$ and $\omega_h = 32\pm0.5 s^{-1}$ respectively. Thermal imaging of the rotation of the top was carried out with the Therm-App TAS19AQ-1000-HZ thermal camera equipped with a LWIR 6.8mm f/1.4 lens. A typical image, representing the temperature field observed under rotation of the heavy top, is supplied in **Figure 2**. A distinct thermal track from the heavy spinning top's neb is present on the surface, whereas in the case of a light top the track is beyond the measurement sensitivity range. The thermal field arising from the spinning of the top is supplied in **Figure 3**. The maximal temperature jump established for the hollow top with a mass of 8.650 grams was 0.3±0.07ºC, whereas the top filled by steel balls ($m$=99.330 g) gave rise to a temperature jump close to one degree centigrade, as shown in **Figure 3**. These results are well expected, due to the fact that the heavier tops led to larger friction forces, resulting in higher heat dissipation. The qualitative observation should be emphasized: when the hollow light top started its rotation on a rough paper when inclined from vertical to $\theta$=26º±0.5°($\theta$ is one of the three Euler angles, see Ref. 13), it continued the inclined spinning until a complete stop (during 30 seconds); whereas the heavy top, inclined at $\theta$= 26º±0.5°was re-oriented vertically after *ca* 8 seconds of rotation. It was observed that the orientation time decreases with the surface roughness increase. It takes 14 seconds to align the heavy top from $\theta$=26º±0.5° to vertical orientation on a polished marble plate, and almost 20 seconds on a glass slide.

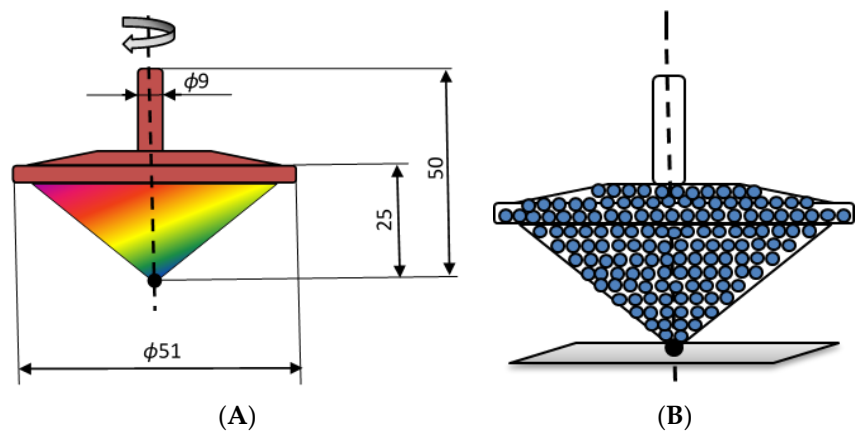

Figure 1.**A**. Scheme of the hollow top used in the investigation. **B**. The top filled with steel balls.

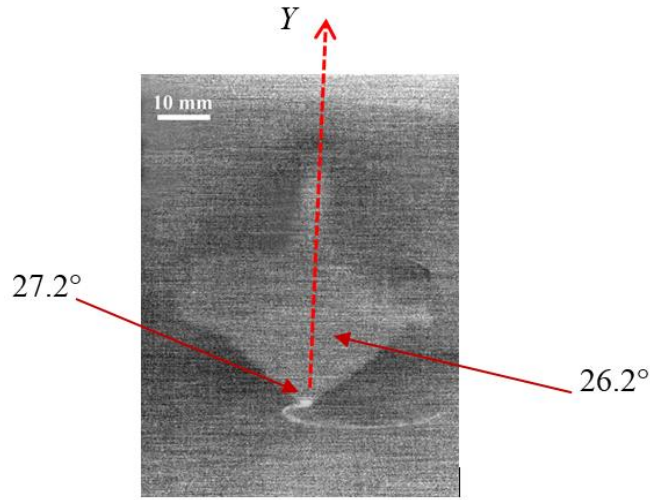

**Figure 2.** Thermal image of the spinning top (the mass $m_f = 99.330g$). Brighter spots correspond to higher temperatures. Bright thermal trace produced by the top on the support is clearly recognized.

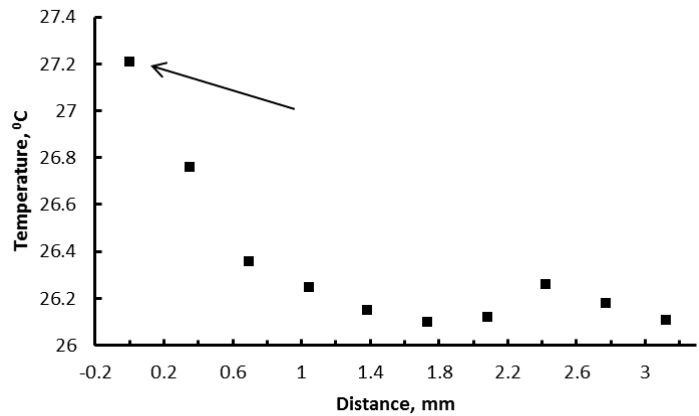

**Figure 3.** The distribution of temperatures along the axis of rotation (axis Y in **Figure 2**) observed for the heavy spinning top ($m_f = 99.330g$).

The second series of experiments was performed with the hollow transparent top depicted in **Figure 1A**, filled by a number of agate (*m*=1.340g *D*=9.7±0.05mm) and steel balls. The number of balls was varied from one to four, as depicted in **Figures 4-5**. The angular velocities of tops at quasi-steady spinning time-spans were $\omega = 50 \pm 0.5 s^{-1}$ for agate balls and $\omega = 80 \pm 0.5 s^{-1}$ for steel ones. The use of the transparent top allowed direct visualization of the location of steel and agate balls within the top. It was observed that spinning ball-filled tops always adopted a symmetric configuration of balls, where the axis of symmetry was the free axis rotation of the system comprised of the top and balls. When a top was filled by a single agate ball, it chose the lowest possible position at the free axis of the top, as depicted in **Figures 4A-B**. A pair and triad of agate balls were located symmetrically relative to the axis of rotation at the equator of the top, as shown **in Figures 4C-D.** Similar results were obtained with the top filled by steel balls, as illustrated in **Figure 5**. It should be emphasized that when the top was filled by four steel balls two symmetrical configurations were possible, as depicted in **Figures 5C-D**; and both of them were experimentally observed. One more experimental observation is noteworthy, namely that when the hollow top was greased from inside by silicone oil

(polydimethilsiloxane, *Mn*=17,000, supplied by Aldrich), the spatial orientation of agate and steel balls within the cavity of the spinner remained the same. This means that viscous friction also promotes symmetrization of the rotating system.

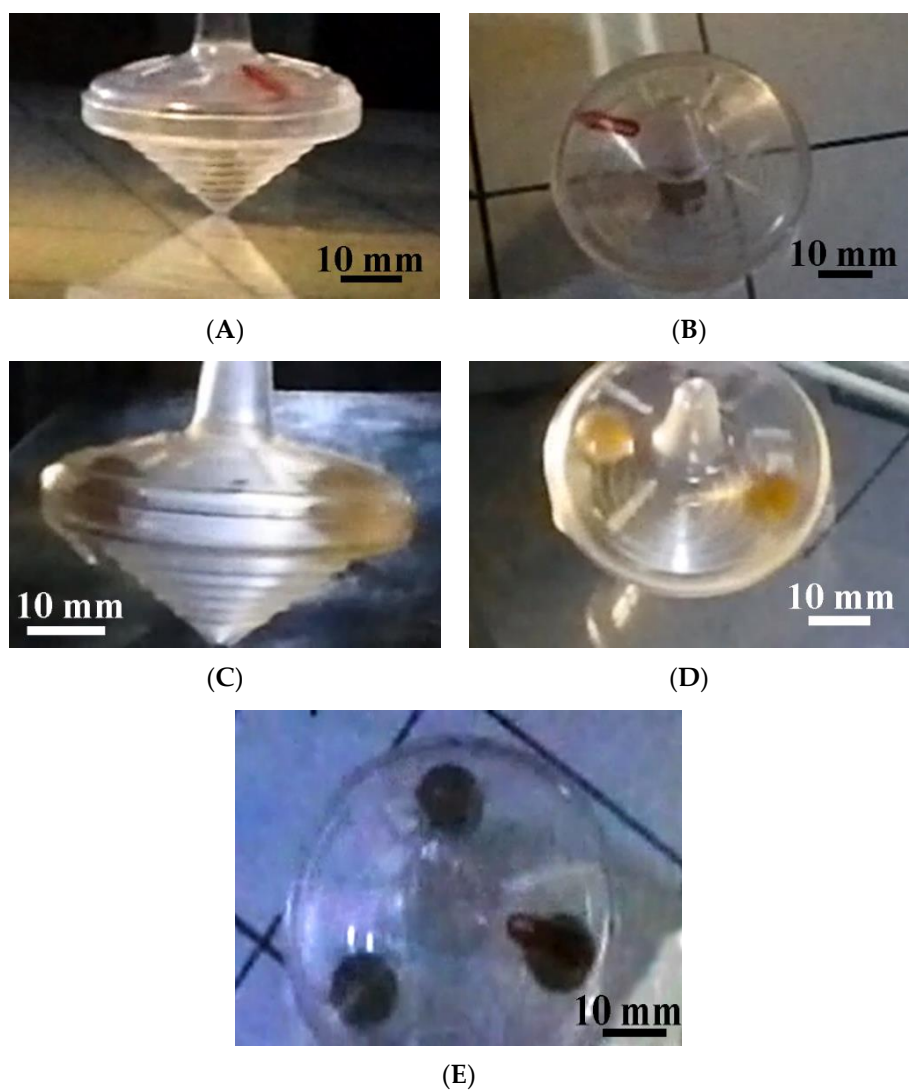

**Figure 4.** Orientation of quasi-steadily rotating transparent tops, filled by agate balls. A-B. Top containing a single agate ball (A. side view, B. plan view). C-D. The same top containing two agate balls. E. Top containing three agate balls.

The third series of experiments was performed with a coin (*m*=6.250g; the diameter *D*=25.9±0.05mm; the thickness *b*=1.7±0.05mm, possessing two holes with the diameter of *d*=3±0.1mm, drilled symmetrically at the distance *l*=11±0.05mm one from another, as depicted in **Figure 6**) rotated on a flat horizontal surface around the vertical axis. In this series of experiments a more complicated rotation was observed: the coin switched the location of holes between two symmetrical orientations admitted by the vertical axis of rotation, as shown in **Figures 6A-B**. The mechanism of this switching is beyond the scope of our paper. However, consider that the orientation of the coin, depicted in **Figure 6A**, corresponds to the maximal moment of inertia *I* of the coin relative to the vertical axis. This kind of rotation was registered for the angular velocity $\omega \cong 8-15 \pm 0.5 s^{-1}$ $\omega$; whereas, the spatial orientation of the coin depicted in **Figure 6B** (that is, the minimal $I$ respective to the same axis) took place for $\omega \cong 17-30 \pm 0.5 s^{-1}$. In both of these experiments, the coin chose the free axis of rotation. The difference between momenta inertia of the coin, oriented as shown in **Figures 6A-B** equals $\Delta I = I_{max} - I_{min} = 2m_0 l^2$, where $m_0$ is the mass of "disk" filling the hole and $l$=5.5 mm is the distance from the symmetry axis to the hole center. The same difference related to the moment inertia of the non-drilled coin $\Delta I / I = 8(m_0 / M) \cdot (l/R)^2 = 8(rl/R^2)^2 \cong 0.02$ where $r$=1.5 mm and $R$=13 mm are the radii of the hole and the coin, respectively.

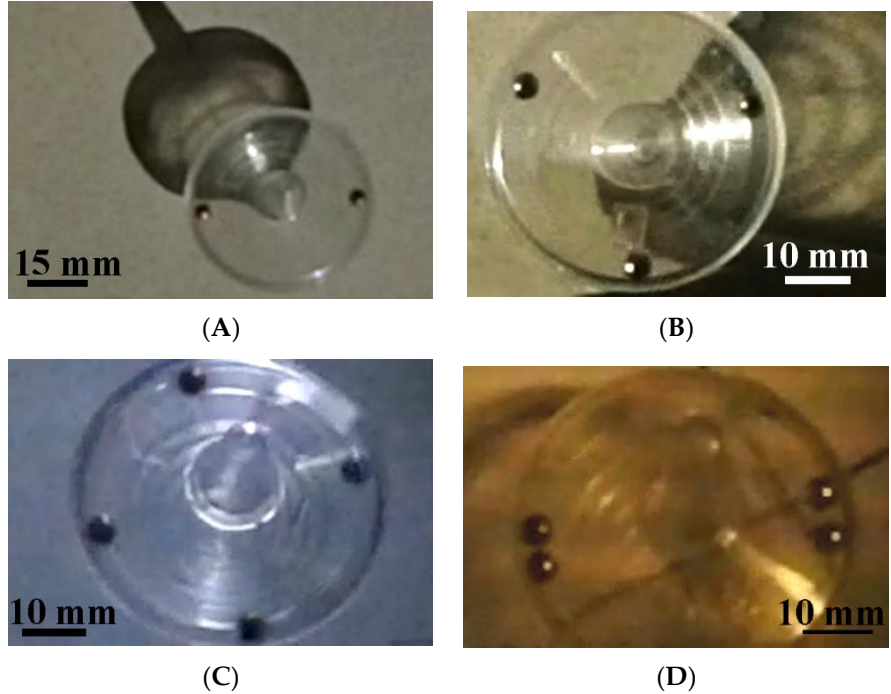

**Figure 5.** Orientation of quasi-steadily rotating transparent tops, filled by steel balls. A. Top containing a pair of two steel balls. B. Top containing a triad of steel balls. C-D. Rotating top containing two pairs of steel balls.

### 3. Estimation of the entropy generation due to the friction-inspired orientation of a rotator.

Consider a rigid rotator possessing a number of free axes of rotation. Recall that free axes of rotation are those keeping their spatial orientation in the absence of external forces [13]. When external forces are absent, a stable rotation occurs around the principal axes of a rigid body, corresponding to maximal and minimal momenta of the rotator [13-14]. However, when friction acts on the rotator, the stable motion is observed around the axis corresponding to the maximal moment of inertia, as demonstrated by the experiments and reported in Sections 2.1-2.3. In solid rotor systems, the effect of friction can be thought of as loss of energy from the rotational mode. However, friction has one more effect: namely, it orients the rotator. It should be stressed that friction force is the only lateral force establishing the symmetrical location of agate and steel balls placed into the transparent spinner, as shown in **Figure 4-5.** This is the friction force that also defines the spatial location of the holes drilled in the rotating coin, depicted in **Figure 6.** The same dry friction establishes the vertical orientation of a "tippe-top" (also known as the Thompson top), as shown in Refs. 15-16.

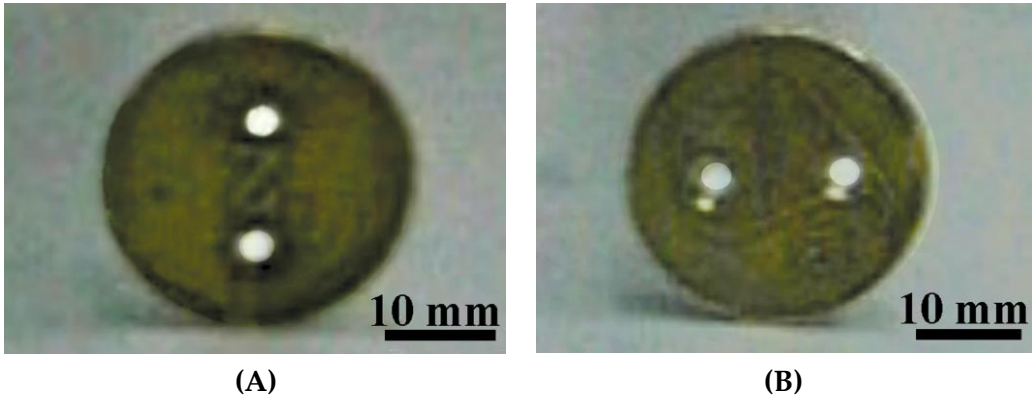

**Figure 6.** Orientation of the rotating coin containing a pair of holes. A. Orientation of holes corresponding to the maximal momentum of inertia of the coin. B. Orientation of holes corresponding to the minimal momentum of inertia of the coin, relative to the vertical axis.

The mechanism of orientation of spinners by friction is discussed in detail in the classical course by Levi Civita and Amaldi (see Ref. 17). Friction force dictates the orientation of the axis of rotation for a variety of tops, and thus decreases the number of rotational degrees of freedom in the system [15-17]. A rectangular parallelepiped falling through air, depicted in **Figure 7**, supplies another illustrative example of orientation of a rotator by air friction forces. Owing to the viscous friction, the body chooses to rotate around the main axis $OO_1$, corresponding to the maximal moment of inertia [18]. This orientation effect leads to the generation of negentropy in a system of rotators.

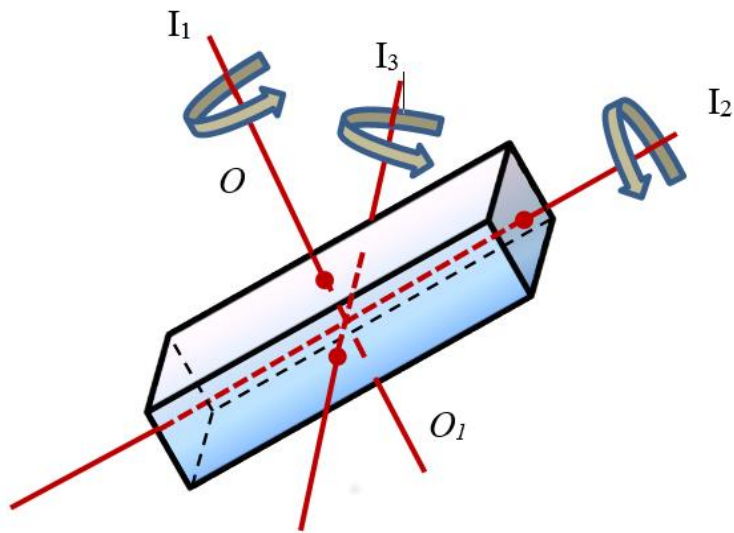

**Figure 7.** Rectangular parallelepiped falling through air is depicted. If inequality $I_1 > I_2 > I_3$ takes place, the falling body will "prefer" rotation around the axis $OO_1$.

Consider a gas of identical rigid rotators influenced by either kind of friction (it seems that such systems were introduced by W. Thomson [19-20]). This gas may be a dust cloud (including the dust plasma [21]), a colloidal suspension built of rigid particles, [22] or a capillary cluster of water droplets [23-24]. If the system is thermally isolated, the total change of entropy with time $\Delta S$ should be necessarily positive, following the Second Law of the thermodynamics:

$$\Delta S = \Delta S_H + \Delta S_{ord} > 0 \qquad (1)$$

where $\Delta S_H = \int \frac{\delta Q}{T} > 0$ is the term describing the generation of entropy in the system of $N$ identical rotators due the heat dissipation caused by friction (either dry or viscous), addressed experimentally in Section 2.1, and illustrated with **Figures 2-3**. $\Delta S_{ord}$ is the negentropy production due to orientation of rotators by the same friction force (again, dry or viscous; see Sections 2.2-2.3 for the experimental study of the orientation of spinners by dry and viscous friction forces). Accurate calculation of the orientational contribution to entropy is an extremely challenging mathematical task [10, 25]. Thus, we propose to estimate the upper boundary of negentropy generation due to eventual elimination of $i$-degrees of freedom by friction (namely $\Delta S_{max}^{ord} < 0$) according to the Sackur-Tetrode formula, supplying the entropy of an ideal gas [26-29]:

$$\Delta S_{max}^{ord} \cong -k_B N \ln\left[\frac{V}{N}\left(\frac{2\pi m k_B T}{h^2}\right)^{i/2}\right] \qquad (2)$$

where $m$ is the mass of the rotator, $V$ is the volume of the system, $k_B$ and $h$ are the Boltzmann and Planck constants respectively, $i$ is the number of degrees of freedom "frozen" by friction; $T$ is the temperature, introduced as the mean kinetic energy of rotators (which is a subtle point in the discussed case [30]; however, for the sake of simplicity, we neglect problems related to the "thermalization" of the gas of rotators).

When friction eliminates two degrees of freedom, we obtain:

$$\Delta S_{max}^{ord} \cong -k_B N \ln\left[\frac{V}{N}\left(\frac{2\pi m k_B T}{h^2}\right)\right] \qquad (3)$$

It should be mentioned that the entropy jump due to loss of rotational/translational degrees of freedom has been intensively discussed in biophysics, regarding bioinformatics of proteins and other complex molecules [31-32]. The maximal orientational entropy given by Eq. 3 corresponds to the minimal possible total entropy $\Delta S_{min}$ observed when all rotators are completely aligned, given by:

$$\Delta S_{min} \cong \int \frac{\delta Q}{T} - k_B N \ln\left[\frac{V}{N}\left(\frac{2\pi m k_B T}{h^2}\right)\right] \qquad (4)$$

When dry friction force acts on rotators, it is reasonable to assume: $\delta Q \neq \delta Q(T)$; thus Eq. 4 may be simplified as follows:

$$\Delta S_{min}(T) \cong \frac{\Delta Q}{T} - k_B N \ln\left[\frac{V}{N}\left(\frac{2\pi m k_B T}{h^2}\right)\right] \qquad (5)$$

where $\Delta Q$ is the total heat dissipated within the process of complete alignment (re-orientation) of rotators. Eq. 1 (arising from the second law of thermodynamics) combined with Eq. 5 yield the non-trivial inequality:

$$\Delta Q > k_B T N \ln\left[\frac{V}{N}\left(\frac{2\pi m k_B T}{h^2}\right)\right] \tag{6}$$

The temperature at which $\Delta S_{\min}(T) = 0$ is of particular interest, hinting at the reversibility of processes occurring in a thermally isolated system of rotators, at the temperature $T^*$, supplying the solution to the transcendental equation:

$$\frac{\Delta Q}{T} = k_B N \ln\left[\frac{V}{N}\left(\frac{2\pi m k_B T}{h^2}\right)\right] \tag{7}$$

It is also instructive to elucidate the temperature dependence of $\Delta S_{\min}$. It is easily seen from Eq. 5 that:

$$\frac{\partial(\Delta S_{\min})}{\partial T} = -\frac{\Delta Q}{T^2} - \frac{k_B N}{T} < 0 \tag{8}$$

It is recognized from Eq. 8 that the minimal change in total entropy, corresponding to the eventual complete alignment of rotators, is decreased with the temperature (i.e. the mean kinetic energy) of rotators. If the friction is viscous, and $\delta Q \sim$ (mean value of $\omega^2$) $T$ takes place [33], we derive:

$$\frac{\partial(\Delta S_{\min})}{\partial T} = -\frac{k_B N}{T} < 0 \tag{9}$$

Again, the minimal jump in total entropy of the system of rotators decreases with temperature. The thermal capacity of a system of $N$ completely oriented spinning tops will obviously be given by:

$$c_v = \frac{1}{2} N k_B \tag{10}$$

which is governed by the single possible degree of freedom.

4. **Conclusions**

Friction forces may produce both entropy and negentropy, as discussed in Refs. 5-9. We address the situation where friction forces promote the orientation of rotating rigid bodies, and thereby eliminate (freeze) a number of their degrees of freedom (the well-known orientation of falling bodies or a tippe-top illustrates an alignment of rotators by dry and viscous frictions [15-18]). We illustrated the friction-inspired orientation of tops by a series of experiments performed with a variety of rotators, including hollow tops filled with rigid balls and rotating drilled coins. In all the studied spinners, dry friction promoted the orientation of a spinner around the free axis of rotation. Thus, the effect of friction results not only in loss of energy, giving rise to heating, but also leads to the orientation of rotators.

A gas of rotators (representing dust plasmas, colloidal and granular systems) influenced by friction was treated theoretically. Friction forces, by aligning rotators, "freeze" a number of degrees of freedom of rotators, and in parallel heat the system (the effect of heating was addressed experimentally). The negative change in the orientational entropy of the gas of rotators was calculated with the Sackur-Tetrode formula. The minimal change in total entropy corresponding to complete alignment of rigid bodies decreases with the temperature, defined as the mean kinetic energy of rotators. The essential effect of the orientational degrees of freedom on the phase equilibrium and motion of systems built from colloidal particles has been reported recently [34-35]. In the broader context, the friction-induced negentropy, owing to elimination of rotational degrees of freedom, is expected to be significant for the phase equilibrium and deformation of granular materials [36-37]. The discussed generation of negentropy by friction may be of primary importance for so-called artificial molecular rotors [38].

**Acknowledgements:** The authors are indebted to Professor Gene Whyman and Mrs. Yelena Bormashenko for their kind help in preparing this manuscript.

Acknowledgement is made to the donors of the Israel Ministry of Absorption for the partial support of the scientific activity of Dr. Mark Frenkel.